\documentclass[aps,natphysics,superscriptaddress,two column,balance,preprintnumbers,pagewise,notitlepage,nobibnotes, nofootinbib]{revtex4-2}

\usepackage[T1]{fontenc}
\usepackage[latin9]{inputenc}
\usepackage{color}
\usepackage{units}
\usepackage{amssymb}
\usepackage{amsmath}
\usepackage{graphicx}
\usepackage{esint}
\usepackage{bm}
\usepackage{natbib}
\usepackage[breaklinks=true,colorlinks=true,urlcolor=blue,
citecolor=blue,linkcolor=blue,bookmarks=false]{hyperref}
\usepackage{babel}
\usepackage{amsfonts}
\usepackage{subfigure}
\usepackage{dcolumn}
\usepackage{xcolor}

\usepackage{acronym}
\newacro{AI}{artificial intelligence}
\newacro{ANN}{artificial neural network}
\newacro{FM}{ferromagnet}
\newacro{HM}{heavy-metal}
\newacro{DW}{domain wall}
\newacro{eNVM}{emerging non-volatile memory}
\newacro{RRAM}{resistive random access memory}
\newacro{PCM}{phase change memory}
\newacro{QNN}{quantized neural network}
\newacro{LLGS}{Landau-Lifshitz-Gilbert-Slonczewski}
\newacro{MTJ}{Magnetic Tunnel Junction}

\newacro{DMI}{Dzyaloshinskii-Moriya interaction}
\newacro{CIP}{current-in-plane}
\newacro{CPP}{current-perpendicular-to-plane}
\newacro{PMA}{perpendicular magnetic anisotropy}
\newacro{AFM}{anti-ferromagnetic}
\newacro{LIF}{leaky-integrate-and-fire}
\newacro{LTD}{long-term depression}
\newacro{LTP}{long-term potentiation}
\newacro{SNN}{spiking neural network}
\newacro{TL}{top layer}
\newacro{BL}{bottom layer}
\newacro{OOMMF}{object-oriented micromagnetic framework}
\newacro{MTJ}{magnetic tunnel junction}

\newacro{FL}{free layer}
\newacro{PL}{pinned layer}
\newacro{SOT}{spin-orbit torque}
\newacro{SHE}{spin-Hall effect}
\newacro{VMM}{vector-matrix multiplication}

\usepackage[margin=0.54in]{geometry}
\usepackage{xcolor}
\usepackage{hyperref}

\begin{document}
	\title{Spintronic Neuromorphic Hardware Using Domain Wall-Based Neurons and Quantized Synapses}
	
	\author{$\mathrm{Sakshi~Kiran~Bandekar^{}}$}
	\affiliation{Department of Electrical and Electronics Engineering, Birla Institute of Technology and Science, Pilani, K K Birla Goa Campus, Goa 403726, India}
	
	\author{Arnab Ganguly}
	\affiliation{Department of Physics and Nanotechnology, SRM Institute of Science and Technology, Kattankulathur, Tamil Nadu 603203, India}
	
	\author{$\mathrm{Debanjan~Polley^{*}}$}
	\thanks{These authors contributed equally to this work and share corresponding authorship.\\ Corresponding author email: debanjan.polley@hyderabad.bits-pilani.ac.in}
	\affiliation{Department of Physics, Birla Institute of Technology and Science, Pilani, Hyderabad Campus, Telangana 500078, India}

	\author{$\mathrm{Debasis~Das^{*}}$}
	\thanks{Corresponding author email: debasis.das@goa.bits-pilani.ac.in}
	\affiliation{Department of Electrical and Electronics Engineering, Birla Institute of Technology and Science, Pilani, K K Birla Goa Campus, Goa 403726, India}

	\begin{abstract}
		\noindent In this work, we simulate the functionality of an artificial neuron and synapse using spin-orbit torque-based spintronic devices and implement a fully connected \ac{ANN}. These neuro-synaptic devices are emulated via transverse domain wall dynamics within a rectangular magnetic nanotrack comprised of \ac{HM}/\ac{FM} heterostructures. Here, the heterostructure is parameterized using Pt-based alloys for the \ac{HM} layer and CoFeB for the \ac{FM} layer. The ReLU activation function of the neuron has been mimicked using the \ac{DW} motion induced by a 3 ns current pulse. 
		The synapse is modeled using current-induced \ac{DW} dynamics with a corrugated \ac{HM}/\ac{FM} nanotrack. 
		Semicircular notches are symmetrically positioned along both edges of the nanotrack to serve as pinning sites.
		By applying 10 ns current pulses of varying current densities, we achieve controlled, stepwise \ac{DW} motion characterized by temporary pauses at consecutive pinning centers.		
		The electrical conductance of the pinned \ac{DW} across various pinning points act as stable synaptic weights for our \ac{ANN}. Furthermore, we observe a threshold-dependent delay effect where each depinning event is influenced by previous ones, successfully mimicking synaptic states and adaptability in neuromorphic systems. The fully connected \ac{ANN} has been modeled using the conventional float32 synaptic weights for the MNIST and Fashion-MNIST datasets with an accuracy of $\sim$ 97\% and $\sim$86\% respectively, which serves as a test bed for our neuromorphic simulations. 
		To implement a sparse and low-memory-footprint \ac{ANN}, we quantize the trained synaptic weights into discrete quantized levels and tested the network, which demonstrate an accuracy of $\sim$95\% and $\sim$62\% for the MNIST and Fashion-MNIST datasets, respectively. 
		Here, we demonstrate that the write and read energies of the three-layer fully connected neural network are on the order of pJ, with an operational latency of 8 ns.
		Although direct quantization impacted performance, fine-tuning the network, fully restored accuracies to near-baseline levels.		 
		These findings highlight the potential of engineered DW pinning-depinning dynamics for scalable, adaptive, and hardware-efficient neuromorphic computing.
	\end{abstract}
	
	\maketitle
	\section{Introduction}
	
	The history of \ac{AI} dates back to the decade of 1940's, led by the pioneering work of Warren McCulloch and Walter Pitts\cite{mcculloch1943logical},  Donald O. Hebb\cite{hebb-organization-of-behavior-1949}. The usage of \ac{AI} has exploded exponentially over the last decade, due to the development of semiconductor fabrication technology and various hardware accelerators\cite{li2016evaluating}.
	Scientists are using \ac{AI} in almost every field spanning from material discovery\cite{hong2020machine}, band structure calculation\cite{lee2016prediction,tarbi2022bandgap}, drug discovery\cite{agatonovic2000basic, baskin2016renaissance}, computer vision\cite{krizhevsky2012imagenet, gopalakrishnan2017deep}, medical imaging\cite{xiao2024convolutional,pan2015brain}, \emph{etc}.
	Inspired by the highly interconnected, parallel processing of biological neurons and the nervous system's ability to learn and adapt, \acfp{ANN} were designed as computational systems that mimic this organization by adjusting weighted connections to analyze complex information and map inputs to outputs.
	However, implementing massive parallelism on traditional hardware is hindered by the von Neumann bottleneck, which requires constant, energy-intensive data transfer.
	Therefore, researchers implement \acp{ANN} using \ac{eNVM} devices such as resistive random access memory \cite{yu2018neuro,yu2021rram,baroni2022low}, phase change memory \cite{burr2015experimental, burr2016recent,sebastian2019computational,wang2024reconfigurable}, spintronics\cite{sengupta2016proposal,ross2023multilayer,chen2023spintronic,kumar2024bimodal} devices.
	Among these various \ac{eNVM} technologies, spintronics-based devices stand out as a superior choice owing to ultrafast dynamics\cite{seifert2016efficient, polley2023picosecond, polley2022progress}, low power consumption\cite{fan2015stt,zhang2014spintronics}, and high endurance\cite{finocchio2021promise}.
	Fundamental spintronic device such as \ac{MTJ} offers two distinct stable resistance states\cite{fong2015spin,mondal2023single} based on the magnetoresistive\cite{parkin2004giant} phenomenon.
	Such devices with binary states have proven to be an excellent choice for implementing digital memory \cite{fong2011bit,he2017exploring,wang2018novel}; however, neuromorphic computing requires memory elements capable of mimicking the inherently analog function of the biological brain.
	Magnetic \acf{DW} is a magnetic texture that separates magnetic domains of opposite magnetization, and its precise manipulation using electric currents or magnetic fields makes DWs promising candidates for high-density memory, logic operations, and adaptive computational systems.
	Such magnetic \ac{DW}-based devices could mimic the analog nature by utilizing the position of the \ac{DW} to encode multiple intermediate resistance states, which directly implement the continuously adjustable synaptic weights.
	Several \ac{DW}-based neuromorphic studies have been carried out to emulate either basic sigmoidal curves or \ac{LIF} neurons.
	For example, Liu et~al.\ \cite{liu2024domain} demonstrated \ac{DW}-driven \ac{MTJ} devices implementing sigmoidal activation functions and multi-state synapses at 508~fJ/operation, where the pinning mechanism relies on selective layer etching.
	In another study, Yang et~al.\ \cite{yang2021integrated} demonstrated an integrated neuromorphic network using artificial spin synapses and spin neurons in a crossbar array configuration with Hall voltage readout, achieving 93\% accuracy on a classification task.
	For \ac{SNN}-based works, Wang et~al.\ \cite{wang2023spintronic} demonstrated spintronic \ac{LIF}-type spiking neurons with self-reset and winner-takes-all functionality using a synthetic antiferromagnet heterostructure, achieving an energy efficiency of 486~fJ/spike and 88.5\% accuracy on the MNIST dataset.
	Apart from that, Lone et~al.\ \cite{lone2025spintronic} addressed \ac{DW}-based \ac{HM}/\ac{FM} heterostructures targeting an \ac{SNN} paradigm with \ac{LIF}-type spiking neurons, achieving $\sim$96\% accuracy.
	However, such implementations primarily focus on spiking paradigms and often rely on external magnetic fields or unconstrained continuous states to maintain operational stability. 	
	Beyond domain walls, other topologically protected magnetic textures, such as magnetic skyrmions~\cite{raj2026reconfigurable} and vortex oscillators~\cite{shreya2023granular}, have been explored for neuromorphic computing.
	While promising, their diffusive motion dynamics present challenges for precise, deterministic state control.
	\acp{DW}, on the other hand, can be deterministically positioned at lithographically defined pinning sites to provide stable, quantized conductance states.
	Despite these advances, several critical challenges, specifically, implementing hardware-friendly ReLU activations for standard \acp{ANN}, establishing a direct link between device-level quantized conductance states and a complete \ac{QNN} training pipeline, and securing synaptic weight stability against stray fields via engineered geometric pinning need to be thoroughly explored. 	
	To address these challenges, in this work, we propose a comprehensive device-to-network framework that leverages \ac{SOT}-driven \ac{DW} motion in \ac{HM}/\ac{FM} nanotracks to emulate both a hardware-friendly ReLU neuron and a stray-field-resilient, quantized artificial synapse for fully connected neural networks.

	A neural network could be trained online, where the training involves direct modification of the synaptic weights within the non-volatile memory hardware itself, a process that is often time-consuming due to the low speed of iterative write/erase cycles and costly due to the high energy consumption per training epoch. Alternatively, the network can undergo offline training (or ex-situ training), where the training is performed using high-performance computing resources, such as GPUs or conventional CPUs, utilizing frameworks like PyTorch. After completion of the training, the final, optimized synaptic weights are transferred to and fixed within the non-volatile memory hardware for the inference stage. This approach significantly reduces the latency and energy costs associated with the long training process on the resource-constrained \ac{eNVM} hardware. Thus, stability of the synaptic weight is of utmost importance when an \ac{ANN} is used for inference on edge devices with an expectation of longer lifetime. However, due to stray fields or demagnetization fields, the position of the \ac{DW} could be displaced\cite{cui2020maximized}, which would hinder the stability of the pretrained synaptic weights and eventually cause a lower inference accuracy. Therefore, the necessity of having a well-controlled synapse that exhibits high stability for holding synaptic weights becomes the central focus for enabling reliable, long-term inference on power-constrained edge devices using \ac{DW}-based neuromorphic hardware. With the ongoing development of spintronics-based fabrication technology, we can surely hope for implementing a complete \ac{DW}-based neuro-synaptic hardware. However, at this early stage, this requires a well-defined model for the \ac{DW}-based neuron and synapse to accurately capture their behavior, followed by a thorough evaluation of the performance of the \ac{ANN}, built using these model-defined components. With this aim in mind, we have designed a magnetic \ac{DW}-based neuron and synapse device on a thin HM/FM nanotrack and performed micromagnetic simulations to obtain their corresponding functionalities. We propose a design for the \ac{DW}-based neuron that mimics the functionality of a simple and hardware-friendly ReLU activation function to bypass the known issues of the dead-neuron problem that occurs in the sigmoid activation function\cite{szandala2020review}. We have added well-defined pinning sites across the nanotrack to ensure the stability of the synaptic weights against any unwanted displacement due to a stray magnetic field. This gives rise to quantized conductance states, which are used to design a fully connected \ac{QNN}. First, a software-based training is performed using PyTorch to train a fully connected neural network, with the custom ReLU-like activation function obtained from our proposed neuron. Then, the trained float32 (FP32) weights are quantized according to the discrete conductance states obtained from the proposed synapse device, and the network is trained again for fine-tuning those quantized weights. The network with the fine-tuned weights shows inference accuracy very similar to the network with FP32 weight counterpart.
	\begin{figure*}[t!]
		\centering
		\includegraphics[width=0.8\textwidth]{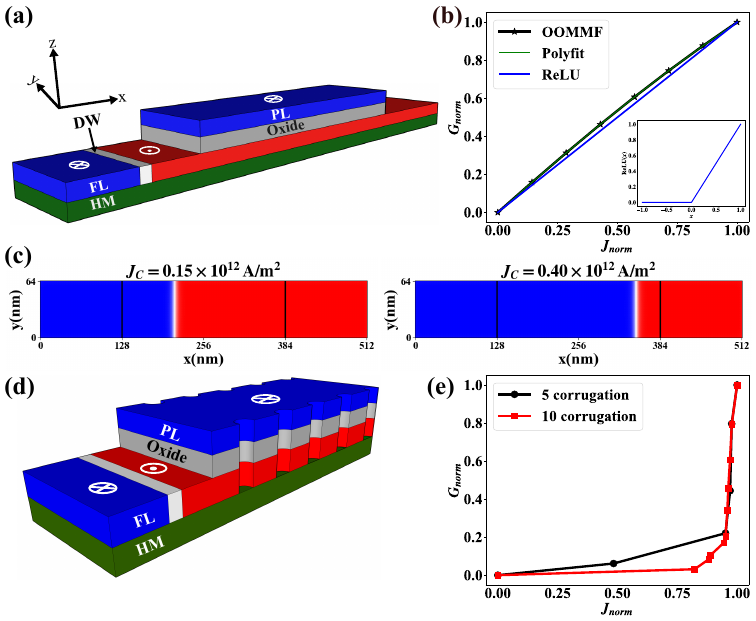}
		\caption{\textbf{Schematic of the proposed spintronic devices to mimic both neuron and synapse.}
			(a) Schematic 3D figure of the proposed neuron device; (b) Plot of normalized conductance vs. current density for the neuron, demonstrating a ReLU-like activation function for the positive input. Inset shows the plot of the actual ReLU activation function for both positive as well as negative input values; (c) Two-dimensional color plot for $m_z$ on the $x\mbox{-}y$ plane of the $FL$ for $J_{c}=0.15 \times 10^{12}$ and $0.40 \times 10^{12}$ $\mathrm{A/m^2}$; (d) 3D schematic of the proposed synapse device using symmetric notch-type corrugation; (e) Plot of normalized conductance vs. current density for the synapse using different numbers of symmetric notches across the nanotrack}.
		\label{Fig:Neuron}
	\end{figure*}

	\section{Micromagnetic simulation details \& Device structure}\label{sec:micomag}
	In this section, we describe the device structure for our proposed artificial neuron and synapse and the details of the micromagnetic framework that has been used for the simulation.
	We simulated both neuron and synapse functionalities by leveraging the temporal dynamics of magnetization (\emph{i.e.}, the \ac{DW} motion) within a HM/FM nanotrack. The micromagnetic simulation has been performed by numerically solving the \ac{LLGS} equation\cite{das2023bilayer} as given below
	\begin{align}
		\begin{split}
			\frac{d\mathbf{m}}{dt}=&\gamma\left(\mathbf{m}\times\mathbf{H}_{eff}\right) + \alpha\left(\mathbf{m}\times\frac{d\mathbf{m}}{dt}\right)\\ &+\gamma\beta\epsilon\left(\mathbf{m}\times\mathbf{m}_p\times\mathbf{m}\right) \mbox{-}~\gamma\beta\epsilon^{\prime}\left(\mathbf{m}\times\mathbf{m}_p\right)
			\label{LLGS}
		\end{split}
	\end{align}
	where $\mathbf{m}$ is the magnetization unit vector, $\mathbf{H}_{eff}$ is the effective magnetic field consisting of uniaxial anisotropy field, demagnetization field, external magnetic field, and the field due to \ac{DMI}.
	$\gamma$ is the gyromagnetic ratio, $\alpha$ is the damping factor, and $\mathbf{m}_p$ is the spin polarization direction of the spin current in the \ac{FM} layer. Apart from these, $\beta=\hbar P J/\left(2\mu_0 e t_{FM}M_S\right)$, and $\epsilon=\frac{P^2\Lambda}{\left(\Lambda^2+1\right)+\left(\Lambda^2-1\right)\left(\mathbf{m}\cdot\mathbf{m}_p\right)}$.
	Details of the parameters are given in Table \ref{Table_param}.
	\begin{table}[t]
		\renewcommand{\arraystretch}{1.3}
		\caption{Device simulation parameters}
		\centering
		\begin{tabular}{l c c c c}
			\hline
			Parameter & Symbol & Value \\
			\hline
			Gyromagnetic ratio & $\gamma$ & 2.211$\times 10^5$ m/(A$\cdot$s)\\
			Reduced Planck's Constant & $\hbar$ & 1.054$\times 10^{-34}$ J$\cdot$s\\
			Free space permeability & $\mu_0$ & 4$\pi \times 10^{-7}~\mathrm{N/A^2}$\\
			Electronic charge & $e$ & 1.602$\times 10^{-19}$ C\\
			FM free layer thickness & $t_{FM}$ & 2 nm\\
			Polarization & $P$ & 0.4\\
			Saturation magnetization & $M_s$ & 1.1$\times10^6$ A/m\\
			Exchange stiffness constant & $A_{ex}$ & 2.0$\times10^{-11}$ J/m\\ 
			DMI constant & $D$ & 1-2 $\mathrm{mJ/m^2}$\\
			Anisotropy constant & $K_u$ & $10^6~\mathrm{J/m^3}$\\
			\hline
		\end{tabular}
		\label{Table_param}
	\end{table}
	The magnetization dynamics is simulated in open-source \ac{OOMMF}\cite{OOMMF} with \ac{DMI} extension module\cite{DMI_extension} and UBERMAG\cite{beg2021ubermag}.
	Fig.~\ref{Fig:Neuron}(a) illustrates a 3d schematic of the proposed artificial neuron, which integrates a HM layer with a \ac{FM} layer, referred to as \ac{FL}. The dimensions of both layers are considered to be 512 nm$\times$64 nm$\times$2 nm. We replicate the functionality of the ReLU activation function using our proposed magnetic \ac{DW}-based artificial neuron device. The \ac{FL} contains two magnetic domains: one with magnetization pointing in the $+z$-direction (red region) and the other in the $\mbox{-}z$-direction (blue region). A magnetic \ac{DW} is shown as the white region which separates these two domains, and its position is encoded as the measure of conductance of the device. When an electric current is applied to the \ac{HM} layer along the $+x$-direction, the spin-Hall effect generates spin-polarized electrons along the $\pm y$-direction and accumulate at the top and bottom surfaces of the \ac{HM} layer. 
	This spin accumulation leads to a spin current flowing in the $+z$-direction, which in turn creates a \ac{SOT}. The \ac{SOT} can alter (depending on the amplitude, direction, and duration of the current pulse) the local magnetization of the \ac{FL}, causing the \ac{DW} to move along the $\mbox{+}x$-direction. The position of the DW is being used to measure the conductance of the device, using an \ac{MTJ}-like detector placed on the \ac{FL}, spanning from $x$ = 128 nm to 384 nm. As illustrated in Fig.~\ref{Fig:Neuron}(a), the detector comprises an oxide layer and a \ac{PL}, with its magnetization fixed along the $\mbox{-}z$-direction.

	The synapse has been modeled using a different HM/FM nanotrack. The FM (and the HM) layers are 1500 nm in length, 150 nm in width, and a thickness of 2 nm. However, unlike our neuron model, we have introduced equally spaced semicircular corrugations (notches), each with a diameter of 20 nm, on both edges (top and bottom) of the nanotrack, as shown in Fig.~\ref{Fig:Neuron}(d), ensuring a balanced pinning potential and preventing \ac{DW} tilt. We have varied the number of corrugations (5 and 10 notches) to get different numbers of synaptic weights for the \ac{ANN}. The DW position is approximated using the average magnetization of the rectangular nanotrack along its length. The magnetization dynamics have been simulated in the presence of a 10 ns square current pulse along the length ($x$-axis) of the nanotrack with varying current density. A DMI energy of 1 $\mathrm{mJ/m^{2}}$ has been considered to mimic the HM/FM interface and deterministically switch the magnetization via symmetry breaking.
	\begin{figure*}[!t]
		\centering
		\includegraphics[width=1.0\textwidth]{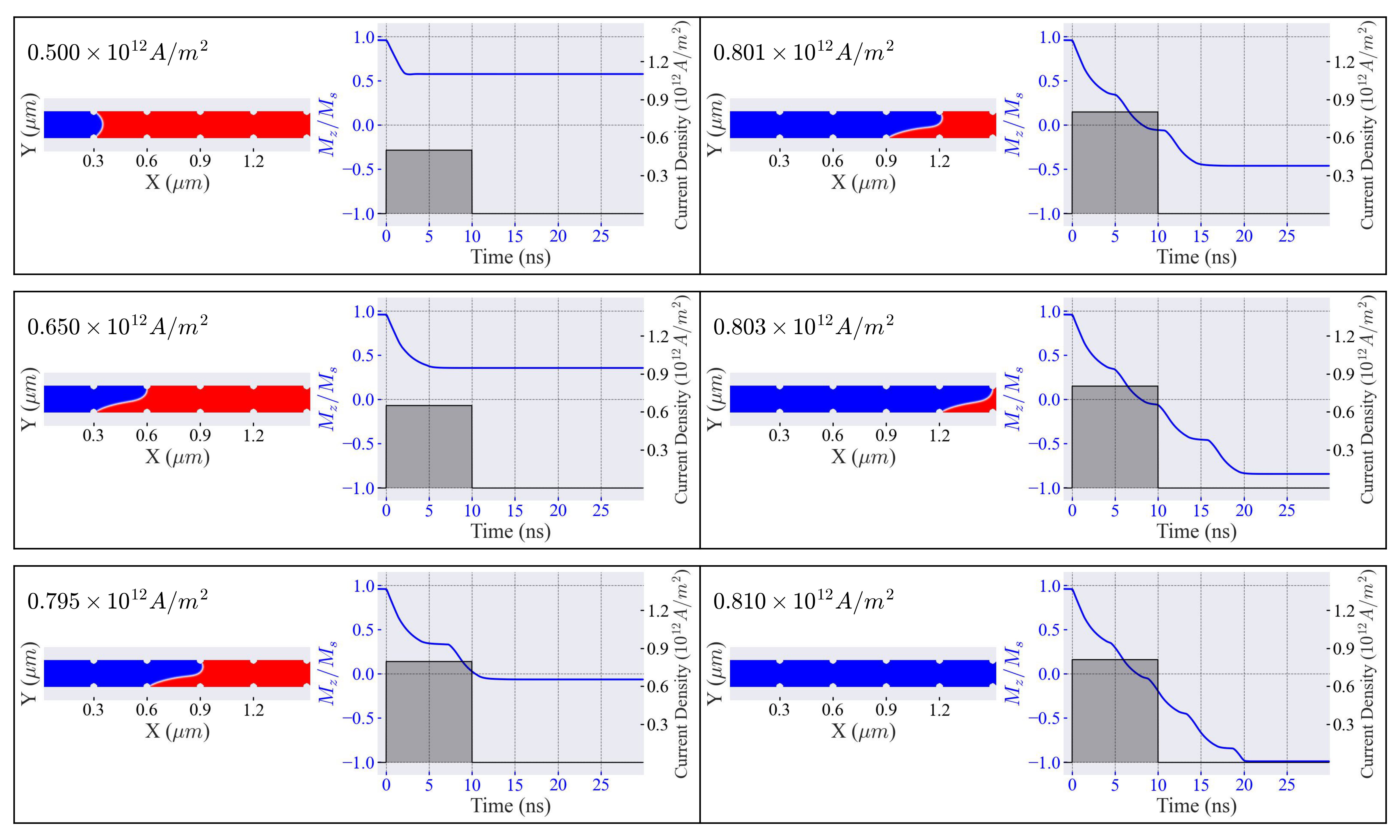}
		\caption{
			\textbf{Domain wall dynamics through the corrugated magnetic nanotrack containing 5 symmetrical notches.}
			The magnetization profile within the corrugated magnetic notches as a function of increasing current density of the 10 ns current pulse. The magnetization profiles are shown on the left, and the corresponding time-resolved magnetization dynamics is shown at right at different current densities. The black-shaded region depicts the 10 ns current pulse with increasing current density. The step-like behaviour suggests the pinning of the domain wall at the engineered equidistant notches on the magnetic nanotrack.
		}
		\label{Fig:Notch}
	\end{figure*}

	\section{Results and Discussion}
	In this section, we discuss the emulation of neuron and synapse functionality using our proposed \ac{DW}-based devices. As mentioned in Sec. \ref{sec:micomag}, the dimension of the \ac{FL} is 512 nm along length, 64 nm along width, with a thickness of 2 nm; we used a mesh size of 1 nm$\times$1 nm $\times$2 nm in our simulation. Initially, the \ac{DW} was outside of the detector, \emph{i.e.}, towards the left side of the nanotrack, and the detector region of the \ac{FL} had the magnetic domain whose magnetization was along the $\mbox{+}z$-direction (Fig. \ref{Fig:Neuron}(a)). At this stage, the conductance is minimal due to the opposite magnetization direction of the \ac{PL} and \ac{FL}.  Conductance across the detector is calculated using the conductance of the \ac{MTJ}-like structure, which is given by\cite{huang2017magnetic}
	\begin{equation}
		G=G_0 \sum_{i}\frac{1+P^2 cos\theta_i}{1+P^2}
		\label{Eq:G}
	\end{equation}
	where, $G_0$ is the conductance when all spins of the \ac{FL} and \ac{PL} of the detector are perfectly parallel to each other. $\theta_i$ is the relative angle between the magnetizations of the $i$th cell of \ac{FL} and the corresponding cell of the \ac{PL} in the detector. With an increase in current density ($J$), the \ac{DW} moves along the $\mbox{+}x$-direction, and the dominance of the $\mbox{-}z$ magnetized domain increases in the detector region. We apply an ideal square current pulse with different current density values from $J$= 0.1 $\times$ $10^{12}$ to 0.45 $\times$ $10^{12}$ $\mathrm{A/m^2}$ with a step of 0.05 $\times$ $10^{12}$ $\mathrm{A/m^2}$, for a span of 3 ns. 
	For a practical case, there would be rise and fall time which will be on the order of a few ps. As the applied current pulse is on the order of 3 ns, which is much larger than the rise and fall time, neglecting these would not affect the dynamics of the \ac{DW}. Hence, for simplicity, we assume the rise and fall time to be 0, for our micromagnetic simulation.
	Fig.~\ref{Fig:Neuron}(b), illustrates the how the normalized conductance ($G_{norm}$) changes with the normalized current density ($J_{norm}$), as shown by the label `OOMMF'. Here, both the parameters are normalized using the formula: 
	\begin{equation}
		f_{norm}=\frac{f-f_{min}}{f_{max}-f_{min}}
		\label{Eq.norm}
	\end{equation}
	where, $f$ is the variable we want to normalize, $f_{max}$ ($f_{min}$) is the maximum (minimum) value of the corresponding variable. The ReLU activation function is given by $f(x)=max(0,x)$, which basically states that for \textit{negative} values of $x$, it remains 0, and for \textit{positive} values, it follows the straight-line formula of $y=x$. The function is shown in the inset of Fig.~\ref{Fig:Neuron}(b). Our device characteristics, \emph{i.e.}, the normalized plot of $G_{norm}$ vs. $J_{norm}$, shows a slight deviation from the ideal ReLU function, as shown in the figure. To account for the deviation, we fit our obtained characteristics with a $4^{th}$ order polynomial, as shown by the green line with a label of `Polyfit'. The negative part of the characteristics is not shown exclusively in the figure, as the negative current density leads to the motion of the \ac{DW} towards the left side, and for every negative current density value, the conductance remains at its minimum value, which translates to zero value according to Eq.(\ref{Eq.norm}). Fig~\ref{Fig:Neuron}(c) shows two-dimensional color plots of the current-induced \ac{DW} motion through the nanotrack for 3-ns square current pulse with current densities $J_C$ = 0.15$\times 10^{12}$ and 0.40$\times 10^{12}$ $\mathrm{A/m^2}$. The two black vertical lines represent the region over which the detector is placed. With increasing current density, the \ac{DW} moves toward the right side, and the $\mbox{-}z$ magnetized domain becomes dominant under the detector region, which leads to increasing conductance as seen in Fig~\ref{Fig:Neuron}(b).
	\begin{figure*}[!t]
		\centering
		\includegraphics[width=1.0\textwidth]{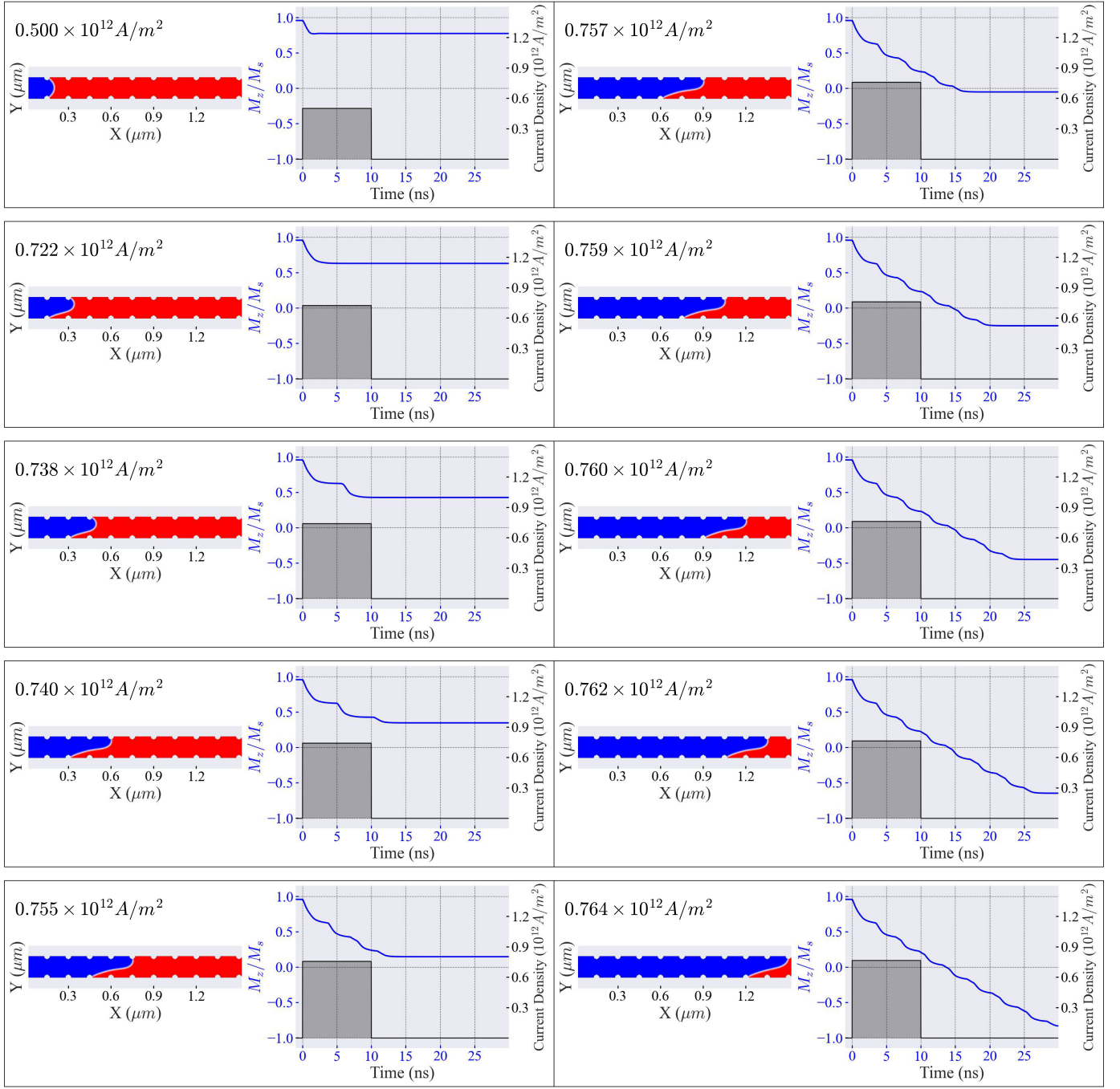}
		\caption{
			\textbf{Domain wall dynamics through the corrugated magnetic nanotrack with 10 symmetrical notches.}
			The magnetization profile within the corrugated magnetic notches as a function of increasing current density of the 10 ns current pulse is shown on the left, and the corresponding magnetization dynamics are shown on the right at different current densities. The black-shaded region depicts the 10 ns current pulse with increasing current density in each case. The step-like behaviour suggests the pinning of the domain wall at the engineered equidistant notches on the magnetic nanotrack.
		}
		\label{Fig:Notch10}
	\end{figure*}
	As mentioned in Sec. \ref{sec:micomag}, the synapse geometry is larger both in length (1500 nm) and width (150 nm); we have used the mesh size as 3 nm$\times$3 nm$\times$2 nm for the micromagnetic simulation. Fig.~\ref{Fig:Notch} illustrates the magnetization profile of the DW dynamics~\cite{ababei2021neuromorphic} in a HM/FL corrugated magnetic nanotrack, due to the application of a 10 ns current pulse with increasing current density. Here, five equally spaced semicircular notches, each with a diameter of 20 nm, act as the DW manipulation centre. The red (blue) color signifies the $z$-component of the magnetization of \ac{FL} pointing up (down). The corresponding time-resolved magnetization dynamics are shown next to the colour profile for each current density, where the domain wall gets pinned at the designated pinning sites. The black-shaded region depicts the 10 ns square current pulse for different current densities. A decreasing magnetization value over time indicates that the DW is moving from left to right, with the slope representing the average velocity of the wall. The first notch has been placed at a distance of 300 nm from the left edge of the magnetic nanotrack, and the spatial distance between each notch (along the length) is 300 nm. Our simulation shows that a current density of $0.5 \times 10^{12}~\mathrm{A/m^{2}}$ is required for the \ac{DW} to stabilize at this first notch. Like the neuron devices, here we also assumed an ideal square current pulse with rise and fall time to be zero. Fig.~\ref{Fig:Notch} shows that it takes $\sim 2~ \mathrm{ns}$ for the domain wall to move to the first notch, translating to a domain wall velocity of $\sim 150~\mathrm{m/s}$.
	The DW dynamics can be understood analytically using the 1D Thiele model for \ac{SOT}-driven \ac{DW} motion~\cite{thiaville2012dynamics,emori2013current}. The \ac{DW} width parameter is given by $\Delta=\sqrt{\frac{A_{ex}}{K_{eff}}}$=9.1 nm (using $K_{eff}=K_u-\left(\mu_0 M_s^2\right)/2$ and the parameters in Table I), which is much smaller than both the notch diameter (20 nm) and the notch spacing (300 nm), confirming well-localised pinning. Below the Walker breakdown, the steady-state \ac{DW} velocity is given by
	\begin{equation}
		v_{DW}=\frac{\gamma_0 \Delta}{\alpha}H_{DL}=\frac{\gamma_0 \Delta}{\alpha} \cdot \frac{\hbar \theta_{SH}}{2 e \mu_0 M_S t_{FM}}\cdot J_C
	\end{equation}
	which predicts $v_{DW}\simeq$315 m/s at $J_C=0.5\times10^{12}~\mathrm{A/m^2}$. This doesn't quite match the simulated results due to the restrictive 1D analytical model without any pinning sites. When the current density is increased to $J_C = 0.650\times 10^{12}~\mathrm{A/m^2}$, the DW bypasses the first notch and gets pinned at the second notch at x = 600 nm. Similarly, at $J_C = 0.795\times 10^{12}~\mathrm{A/m^2}$, $0.801\times 10^{12}~\mathrm{A/m^2}$, and $0.803\times 10^{12}~\mathrm{A/m^2}$, the DW stabilizes at the consecutive notches, located at $x = 900$ nm, 1200 nm, and 1500 nm, respectively. Finally, at $J_C = 0.810\times 10^{12}~\mathrm{A/m^2}$, the \ac{DW} bypasses all notches and traverses the entire nanotrack, and it switches the whole magnetization of the nanotrack from +$z$ to -$z$ direction. The full switching takes $\sim$20 ns using the required current density, which translates to a reduced average domain wall velocity of $\sim 75$ m/s. Thus, the \ac{DW} motion has been engineered by placing the notches in the desired locations. A similar magnetization profile and time-resolved magnetization dynamics have also been shown for a nanotrack containing 10 symmetrical notches in Fig.~\ref{Fig:Notch10}. The pinning of the domain wall at the first notch in the 10-notch system starts at the same current density ($J_{c}=0.500 \times 10^{12}~ \mathrm{A/m^2}$) as for the 5-notch system. However, it takes considerable lesser current ($J_{C}=0.770 \times 10^{12}~ \mathrm{A/m^2}$) to fully switch the 10-notch system (not shown here).
	
	In the magnetic nanotrack with equally spaced, identical notches, each notch serves as a pinning centre by creating a localized energy barrier\cite{noh2012effects}. The notches are lithographically defined geometric corrugations, not crystallographic defects; the DW pinning arises from the local reduction in nanotrack width, which creates a potential energy minimum at each notch center. For the \ac{DW} to move past a notch, its energy, driven by current, must overcome this barrier. The event of depinning, as well as the movement of the \ac{DW} under an applied current, involves energy dissipation through magnetic damping. As a result, the effective potential barrier is increased for each successive notch, requiring larger current densities to overcome the pinning sites. Additionally, the \ac{DW} accumulates elastic strain while pinned, which helps it depin and settle at the next site as the current increases. Each depinning event reconfigures the \acp{DW} chiral structure corresponding energy, making it more susceptible to pinning at the following notch. Notably, the magnetization decay features a step-like pattern, indicating temporary halts of the \ac{DW} motion at each pinning site. The step-like displacement and selective pinning of the DW in response to the current closely parallels neuromorphic computing, mimicking the gradual and quantized changes of synaptic weights in biological synapses. Each pinning site acts as a meta state, allowing incremental adjustments equivalent to synaptic weight updates. This controlled, stepwise depinning is also ideal for implementing gradual learning, where \acp{DW} held at notches retain the conductance state until a higher current prompts movement.
	Without notch-type corrugations, the DW movements are nearly linear, controlled both by the applied current density and DMI energy, which determines the DW velocity.
	The corrugations provide us another knob to control the DW movements and enable lithographic control of synaptic weight values. Other methods for DW pinning include using interlayer exchange coupling of multiple MTJs or by introducing defects in the shape or anisotropy of the FL. However, notches were chosen as the preferred pinning method because their controllable positions guarantee update linearity, which cannot readily be obtained using randomly distributed defects. In addition, notches are less complex to lithographically define than interlayer exchange and can be more easily scaled. Similar to the neuron conductance detector, the DW position is converted to a synapse conductance using a typical MTJ structure, with a parallel resistance of $10~\mathrm{k\Omega}$ and antiparallel resistance of $20~\mathrm{k\Omega}$. Depending on the position of the DW, the resistance changes linearly based on the magnetization of the FL.
	
	To quantify the hardware efficiency of the proposed devices, we estimate the device-level matrices, such as write energy, read energy, latency and device area assuming a Pt alloy-based \ac{HM} layer, where we consider the resistivity, $\rho_{\mathrm{HM}}=80~\mu m\cdot cm$, and thickness $t_{\mathrm{HM}}$=2 nm. As the length and width of the neuron device are 512 nm and 64 nm, respectively, hence the volume of the \ac{HM} is $V_{\mathrm{HM}}=6.55\times 10^{-23}~\mathrm{nm^3}$. Since the neuron is driven by the ideal square current pulse of pulsewidth $\tau$=3 ns, the energy consumption is given by ~\cite{das2023bilayer}
	\begin{equation}
		E_{\mathrm{write}}^{\mathrm{neuron}}=J_C^2 \rho_{\mathrm{HM}} V_{\mathrm{HM}}~ \tau
		\label{Eq:Energy}
	\end{equation}
	where $J_C$ is the applied current density, which lies in the range of $0.1\times10^{12}~\mathrm{A/m^2}$ to $0.45\times10^{12}~\mathrm{A/m^2}$ for our proposed neuron device. Thus, from Eq.(\ref{Eq:Energy}), we obtain the energy consumption for the neuron write operation ranges from 1.57 fJ to 31.8 fJ corresponding to the range of $J_C$. As our proposed synapse has the dimension of length $L$=1500 nm, and width $W$=150 nm, so a \ac{HM} with thickness $t_{\mathrm{HM}}$=2 nm, leads to volume $V_{\mathrm{HM}}=4.5\times 10^{-22}~\mathrm{m^3}$. Since we applied a current density ranging from $J_C=0.5\times10^{12}~\mathrm{A/m^2}$ to $0.81\times10^{12}~\mathrm{A/m^2}$ for a pulse width of 10 ns, Eq.(\ref{Eq:Energy}) estimates an energy consumption ranging from 900 fJ to 2.36 pJ per synaptic write operation.
	In this work, we employ an ex-situ (offline) training scheme and use the trained network for inference only; we should focus on the read operation of the synapse, instead of write. For reading the synaptic states, we use a read voltage of $V_{\mathrm{read}}$=100 mV, and a read time of $\tau_{\mathrm{read}}$=1 ns. The conductance depends on the physical location of the pinned \ac{DW}. From Fig. \ref{Fig:Neuron}(d), we see the magnetization of \ac{PL} is along -$z$-direction. Thus, maximum (minimum) conductance of the synapse is obtained when entire magnetization of the \ac{FL} is parallel (anti-parallel) to the \ac{PL}, \emph{i.e.}, along the -$z$-direction (+$z$-direction). From the micromagnetic simulation, we obtain the magnetization profile corresponding to all the applied current densities, and we use Eq.(\ref{Eq:G}) to obtain conductance ($G$) and finally scaled in a normalized scale ($G_{norm}$) using Eq.(\ref{Eq.norm}). Consider the obtained minimum and maximum conductances are $G_{\mathrm{min}}$ and $G_{\mathrm{max}}$, which would represent the resistance values as $R_{AP}$ and $R_P$, respectively. Using these informations and from Eq.(\ref{Eq.norm}), it can be easily proven that resistance $R_i$, corresponding to a synapse having normalized conductance $G_{\mathrm{norm}_i}$, can be expressed as,
	\begin{equation}
		R_i=\frac{R_{AP} R_P}{R_P + G_{\mathrm{norm}_i}\left(R_{AP}-R_P\right)}.
	\end{equation}
	Hence, the synapses demonstrate a resistance value which lies between $R_{AP}$ and $R_P$. Next, we use
	\begin{equation}
		E_{\mathrm{read}}^i=\frac{V_{\mathrm{read}}^2}{R_i}\times \tau_{\mathrm{read}}
	\end{equation}
	to estimate the energy consumption for any synapse, corresponding to the read operation. We have two different sets of synaptic devices, with 5 and 10 corrugations, and we have used MNIST and Fashion-MNIST to train our neural networks. Considering two layers of synaptic crossbar for Input-to-Hidden and Hidden-to-Output, we obtained total energy consumption per inference image, as shown in Table~\ref{Table:Read_Energy}.
	\begin{table}[t]
		\renewcommand{\arraystretch}{1.3}
		\caption{Synapse Read Energy}
		\centering
		\begin{tabular}{|c|c|c|}
			\hline
			& MNIST & Fashion-MNIST \\
			\hline
			5 corrugation & 59.68 pJ & 60.85 pJ \\
			\hline
			10 corrugation & 59.94 pJ & 61.12 pJ\\
			\hline 
		\end{tabular} 
		\label{Table:Read_Energy}
	\end{table}
	Considering the $L\times W$, the area footprints for neurons and synapses are obtained as 0.033 $\mathrm{\mu m^2}$ and 0.225 $\mathrm{\mu m^2}$, respectively. As our \ac{ANN} architecture is 784-128-10, thus considering all the neurons and synapses, we obtain the device area footprint $\sim 0.023~\mathrm{mm^2}$. We use a square current pulse of pulsewidth 3 ns for the neuron operation and 1 ns for synaptic reading, per inference image. Hence, for two layers, latency would be 8 ns.
	\begin{figure*}[t!]
		\centering
		\includegraphics[width=1.0\textwidth]{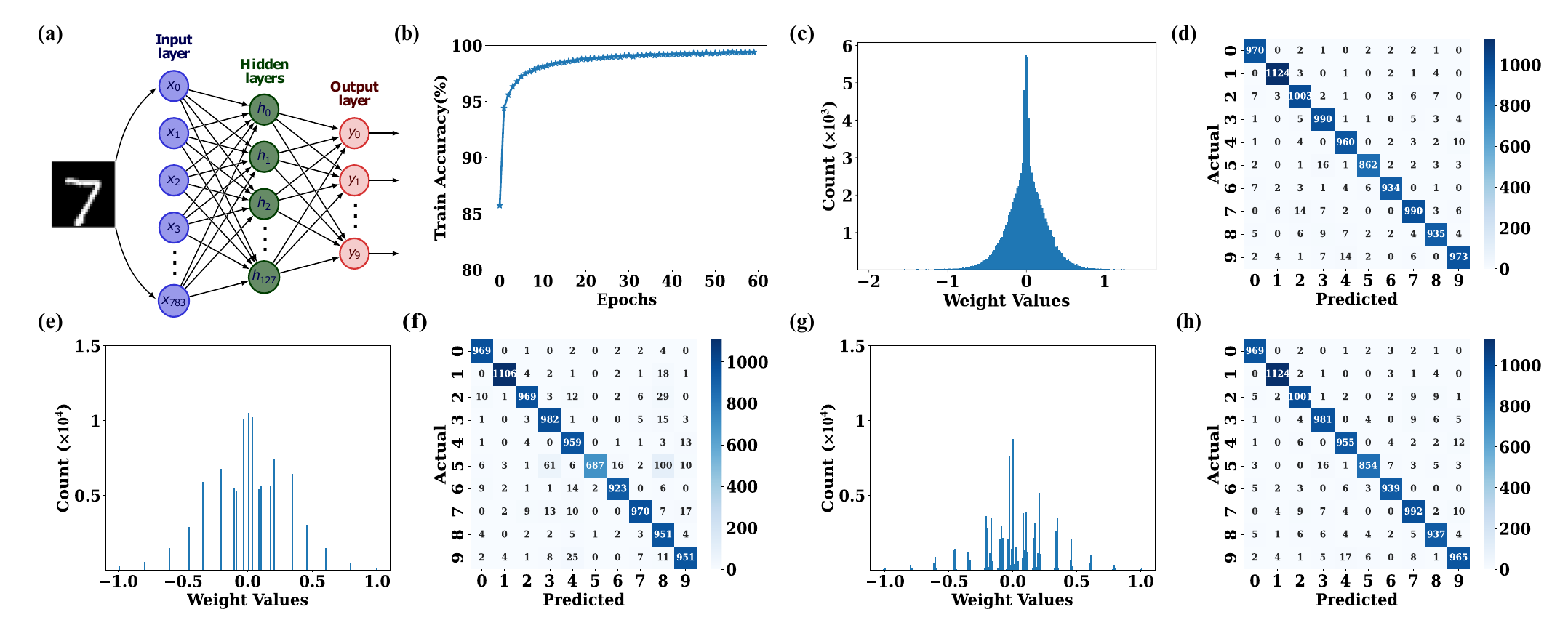}
		\caption{(a) Schematic of fully connected Neural Network for identifying MNIST dataset images, (b) plot of Training accuracy with epochs for FP32 weights, (c) FP32 synaptic Weight distribution from input to hidden layer, (d) Confusion Matrix for FP32 synaptic weights, (e) Synaptic weight distribution of input to hidden layer for pure quantized weights, (f) Confusion Matrix of pure quantized synaptic weights, (g) Synaptic weight distribution for the input to hidden layer fine tuning of the synaptic weights, (h) Confusion Matrix for the fine tuned synaptic weights.}
		\label{fig:mnist_1}
	\end{figure*}
	\begin{figure*}[t!]
		\centering
		\includegraphics[width=1.0\textwidth]{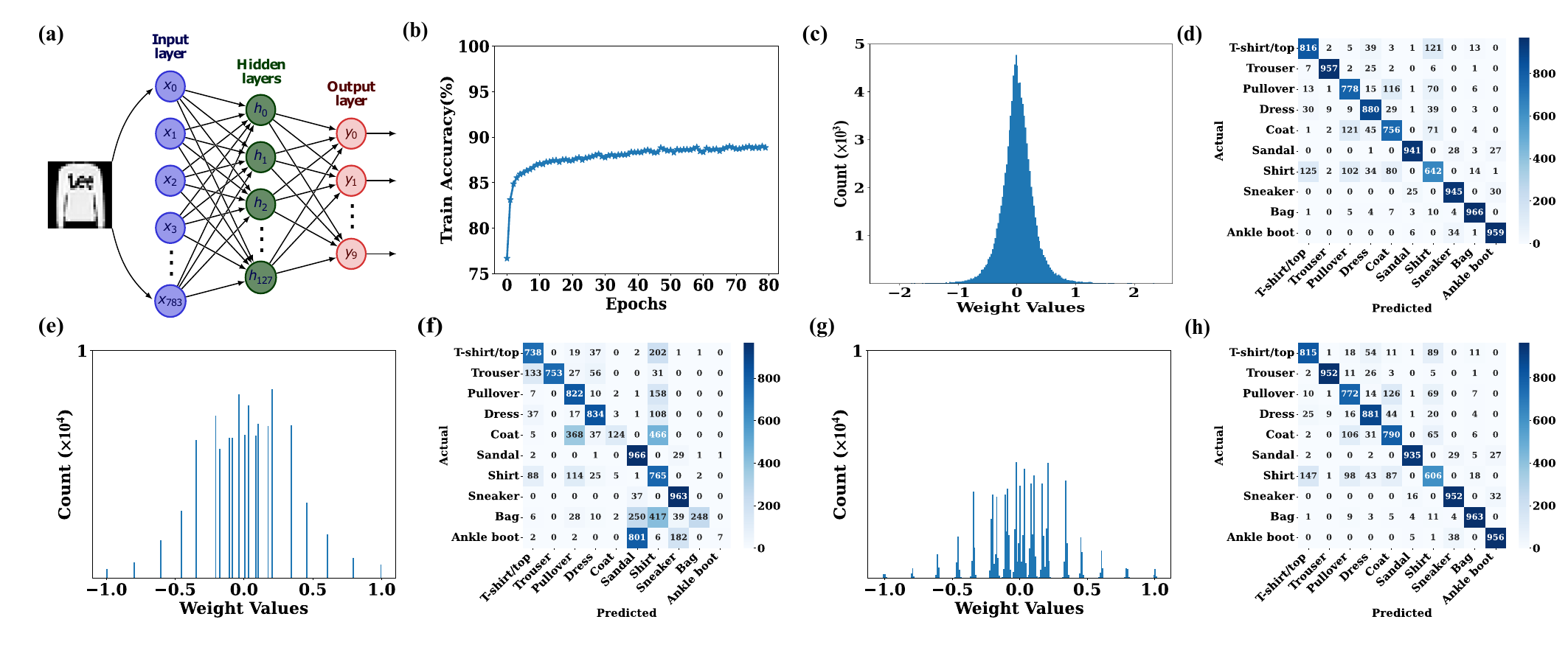}
		\caption{(a) Schematic of fully connected Neural Network for identifying Fashion-MNIST dataset images, (b) plot of Training accuracy with epochs for FP32 weights, (c) FP32 synaptic Weight distribution from input to hidden layer, (d) Confusion Matrix for FP32 synaptic weights, (e) Synaptic weight distribution of input to hidden layer for pure quantized weights, (f) Confusion Matrix of pure quantized synaptic weights, (g) Synaptic weight distribution for the input to hidden layer fine tuning of the synaptic weights, (h) Confusion Matrix for the fine tuned synaptic weights.}
		\label{fig:fashionmnist_1}
	\end{figure*}

	\section{Neural Network Integration}
	In this section, we describe how our proposed devices can be utilized to build a fully-connected \ac{ANN}.
	We train a neural network using two benchmark datasets, the MNIST handwritten digit dataset and the Fashion-MNIST dataset.
	Both datasets contain 60000 training and 10000 test grayscale images with a size of 28$\times$28 pixels.
	Here, we use a three-layer fully connected neural network with 784 nodes for the input layer, 128 nodes for the hidden layer, and 10 nodes for the output layer.
	A schematic of the neural network is shown in Fig.~\ref{fig:mnist_1}(a) and \ref{fig:fashionmnist_1}(a) for the MNIST handwritten digit dataset and the Fashion-MNIST dataset, respectively.
	The activation function of the neurons is implemented using the ReLU-like behavior that is obtained from our proposed neuron device (Fig.~\ref{Fig:Neuron}(b)).
	
	First, we train the fully connected neural network using PyTorch, a package used to build and train machine learning models, where we utilize the default float32 (FP32) weights for the model.
	A distribution of the trained weights for the input to the hidden layer is shown in Fig.~\ref{fig:mnist_1}(c) and Fig.~\ref{fig:fashionmnist_1}(c) for the MNIST digit dataset and the Fashion-MNIST dataset, respectively.
	The simulation results show a test accuracy of $\sim$97\% for the MNIST and $\sim$86\% for the Fashion-MNIST dataset.
	The detailed test accuracy for different variants of our proposed synaptic device is provided in Table~\ref{tab:Table_1}, which shows a negligible variation between these different variants for any particular dataset.
	\begin{table}[h!]
		\centering
		\renewcommand{\arraystretch}{1.0}
		\caption{Result Summary for the test accuracy (in \%)}
		\begin{tabular}{|c|c|c|c|c|}
			\hline
			Conductance Levels & Dataset & FP32 & Quantized & Fine Tune \\
			\hline
			10 corrugation
			& MNIST & 97.41 & 94.67 & 97.17 \\
			& FashionMNIST & 86.40 & 62.20 & 86.22 \\
			\hline
			5 corrugation
			& MNIST & 97.41 & 93.84 & 96.91 \\
			& FashionMNIST & 86.40 & 61.36 & 86.31 \\
			\hline
		\end{tabular}
		\label{tab:Table_1}
	\end{table}
	The key difference between the test accuracies is seen for the MNIST and Fashion-MNIST datasets, which arises due to the complex nature of the Fashion-MNIST dataset with respect to the MNIST one.
	This difference can be understood in depth with the help of a confusion matrix, as shown in Fig.~\ref{fig:mnist_1}(d) and Fig.~\ref{fig:fashionmnist_1}(d), for the MNIST and Fashion-MNIST datasets, respectively.
	As the MNIST images are quite distinct from each other, there is a slight mismatch between the actual and predicted labels.
	On the other hand, the Fashion-MNIST dataset contains many similar-looking ambiguous images with different actual labels, leading to wrong predictions. For example, a key ambiguity comes from the images with labels like `Shirt', `T-shirt/top', `Pullover', which is clearly visible from the number of wrong predictions in the confusion matrix of Fig~\ref{fig:fashionmnist_1}(d), leading to relatively lower test accuracy for the Fashion-MNIST dataset.
	\begin{figure*}[t!]
		\centering
		\includegraphics[width=0.9\linewidth]{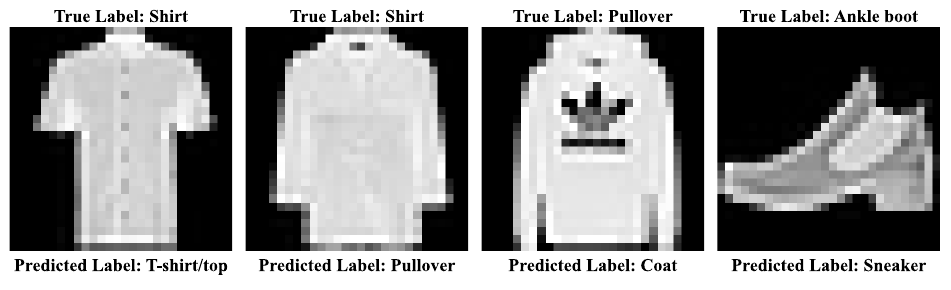}
		\caption{Misinterpreted results of the Fashion-MNIST dataset for FP32 weights.}
		\label{fig:fmnist_fp32_misinterpreted}
	\end{figure*}
	As an example, a few misinterpreted results of the Fashion-MNIST dataset for the FP32 weights are shown in Fig.~\ref{fig:fmnist_fp32_misinterpreted}, which clearly demonstrates the ambiguous nature of the test images. 
	
	To incorporate the effects of distinct synaptic weights of our proposed devices, we quantize the trained FP32 weights into discrete normalized conductance values/levels, mirroring the synaptic states of the DW-based synapse device.
	The obtained conductances from micromagnetic simulations are normalized using Eq.(\ref{Eq.norm}).
	Assume the $i^{th}$-level conductance state is given by $q_i$, where $q_i\in [0,1]$, and the midpoint between two consecutive discrete weights is given by $m_i=\frac{1}{2}\left(q_i + q_{i+1}\right)$, which serve as decision boundaries. As the FP32 weights have magnitude, as well as polarity, they can be separated as $s = \mathrm{sign}(w)$ and $a = \vert w\vert$. The magnitude of the weights can be snapped to the nearest quantized bins via
	\begin{equation}
		Q(a) =
		\begin{cases}
			q_0     & a \leq m_0, \\
			q_k     & m_{k-1} < a \leq m_k, \quad 1 \leq k \leq n-2, \\
			q_{n-1} & a > m_{n-2},
		\end{cases}
		\label{eq:quant}
	\end{equation}
	and the sign is restored to give the final quantized weight $w_q = s \times Q(a)$.
	The distribution of purely quantized synaptic weights connecting the input and hidden layer for the MNIST and Fashion-MNIST datasets are shown in Fig.~\ref{fig:mnist_1}(e) and Fig.~\ref{fig:fashionmnist_1}(e), respectively, for the 10-notch device.
	A similar quantization can also be performed with the conductance states obtained from 5-notch devices, which we do not show here.
	
	The synapses connect the neurons of one layer to those of the next in a fully connected topology. The primary role of each synapse is to scale the incoming signal (current) to a neuron from all previous layer neurons based on its assigned synaptic weight. Consequently, these devices can be arranged in crossbar architecture to perform energy-efficient \ac{VMM} operations~\cite{kumar2024bimodal,jung2022crossbar}. As seen in the trained weight distributions in Fig.~\ref{fig:mnist_1}(c) and Fig.~\ref{fig:fashionmnist_1}(c), the synaptic weights take both positive and negative values. In our architecture, each synaptic weight is represented by the electrical conductance of the proposed \ac{DW}-based synapse (Fig.~\ref{Fig:Neuron}(d)), which is governed by the physical position of the \ac{DW} along the nanotrack (Fig.~\ref{Fig:Notch} and Fig.~\ref{Fig:Notch10}). This \ac{DW} position is converted to a conductance value by applying a potential difference between the \ac{PL} and the right terminal of the \ac{FL}, as expressed numerically by Eq.(\ref{Eq:G}). To realize differential (signed) synaptic weights, the polarity of this applied potential difference is altered. Positive (negative) weights are achieved by applying a higher (lower) potential to the \ac{PL} relative to the \ac{FL}. Since the net synaptic output is computed via \ac{VMM} of the conductance matrix and the input voltage vector, the total current $I$ reaching a neuron is given by
	\begin{equation}
		I=\sum\limits_iG^{+}_i\cdot \Delta V + \sum\limits_j G^{-}_j\cdot \left(-\Delta V\right)
	\end{equation}
	Here, $G^{+}$ and $G^{-}$ represent the conductances corresponding to the positive and negative weights, respectively, and $\Delta V$ is the magnitude of the potential difference applied between the \ac{PL} and the right terminal of the \ac{FL}.
	
	From Fig.~\ref{fig:mnist_1}(e) and Fig.~\ref{fig:fashionmnist_1}(e), we notice that the weights are very sparse and only present at the given synaptic weight levels.
	The corresponding test accuracy for the MNIST dataset is $\sim$94\%; however, a drastic change is noticeable for the Fashion-MNIST dataset, where the test accuracy has degraded to $\sim$62\%.
	The degradation of the test accuracy is due to synaptic weight distribution at purely quantized levels, instead of the entire weight range.
	The degradation in the test accuracy can also be seen by comparing the confusion matrix of Fig.~\ref{fig:fashionmnist_1}(d) and (f), corresponding to FP32 and pure quantized weights.
	Lower values of the diagonal elements of the confusion matrix corresponding to pure quantized weights (Fig.~\ref{fig:fashionmnist_1}(f)) compared to FP32 weights (Fig.~\ref{fig:fashionmnist_1}(d)) reflect this degradation.
	The drastic change corresponding to the Fashion-MNIST dataset is due to the complex and ambiguous nature of the images as discussed earlier.
	Due to the ambiguity among inter-class images, there is a noticeable mismatch in the off-diagonal elements of the confusion matrix in Fig.~\ref{fig:fashionmnist_1}(f).
	For the fine-tuning, we retrain the model with pure quantized weights.
	The weight distributions of the fine-tuned model are shown in Fig.~\ref{fig:mnist_1}(g) and Fig.~\ref{fig:fashionmnist_1}(g) for the MNIST and Fashion-MNIST datasets, respectively.
	This shows a narrow distribution around the discrete quantized labels.
	Such distributions can naturally occur in synaptic weights derived from experimentally fabricated devices, where inherent conductance variations typically fluctuate around a mean value.
	Table \ref{tab:Table_1} shows that the fine-tuned model improves the accuracy, and the test accuracy becomes similar to the FP32 weights.
	Comparison of the confusion matrix between the fine-tuned model and the pure quantized model reflects the improvement of the test accuracy.
	The complete results for the 5-notch and 10-notch systems are summarized in Table \ref{tab:Table_1}.
	These data demonstrate that while the system initially suffers from a degradation in test accuracy, subsequent fine-tuning effectively recovers the performance, bringing the accuracy closer to its baseline (FP32) value.
	
	\section{Conclusions}
	
	We have presented a robust framework for simulating and implementing a fully connected \ac{ANN} utilizing SOT-driven spintronic devices. Using transverse \ac{DW} dynamics in a HM/FM nanotrack, we have successfully emulated the critical functionalities of both artificial neurons and synapses. Our approach bridges the gap between fundamental \ac{DW} driven magnetization switching and neuromorphic computing architectures, offering a viable pathway toward highly efficient, brain-inspired computational hardware. At the device level, our micromagnetic simulations establish that the ReLU activation function of a neuron can be accurately mimicked through \ac{DW} motion triggered by a 3 ns current pulse in a HM/FM nanotrack. More critically, we engineered the synaptic behavior by introducing semicircular notch-type corrugations, symmetrically located along another HM/FM nanotrack. Under the influence of 10 ns current pulse with varying current densities, these geometric constraints effectively serve as discrete pinning sites for the domain wall. This configuration results in a highly controlled, step-like \ac{DW} motion characterized by temporary pauses at each pinning center. We have demonstrated that the electrical conductance of the \ac{DW} at these specific pinned locations provides stable, distinct states that seamlessly translate into programmable synaptic weights for the \ac{ANN}. Furthermore, our investigation revealed a crucial threshold-dependent delay effect during the depinning process.
	Because each depinning event is intrinsically influenced by the history of previous states, the device inherently mirrors the synaptic states found in biological neural systems.
	To validate the system-level performance of our proposed neuro-synaptic devices, we evaluated the fully connected \ac{ANN} using standard FP32 synaptic weights. Testing on the benchmark MNIST and Fashion-MNIST datasets yielded impressive classification accuracies of approximately $\sim$97\% and $\sim$86\%, respectively. 
	At the system level, our proposed neuro-synaptic architecture demonstrates ultra-low power consumption, with read and write energies on the order of picojoules, a compact device footprint of $\sim$0.023 $\mathrm{mm^2}$, and an operational latency of 8 ns for the three-layer fully connected network.
	These baseline results confirm that the physical dynamics of \ac{DW} pinning and depinning can be reliably mapped to complex algorithmic requirements without a fundamental loss of computational fidelity. 
	While this study validates the fundamental device-to-network framework using a fully connected topology, extending this hardware-aware approach to deeper architectures (e.g., convolutional and residual networks) and larger-scale datasets remains an important direction for future system-level exploration. Such extensions will further clarify the scalability and energy efficiency of DW-based spintronic accelerators in complex edge-AI environments.

	\section*{Acknowledgements}
	\noindent This work was supported by the BITS New Faculty Seed Grant (NFSG). Financial support for D.D. and D.P. was provided under grant references N4/24/1005 and N4/24/1018, respectively.
	
	\section*{References}
	\bibliography{Bibliography_Debasis}

@article{mcculloch1943logical,
	title={A logical calculus of the ideas immanent in nervous activity},
	author={McCulloch, Warren S and Pitts, Walter},
	journal={The bulletin of mathematical biophysics},
	volume={5},
	number={4},
	pages={115--133},
	year={1943},
	publisher={Springer}
}

@incollection{szandala2020review,
  title={Review and comparison of commonly used activation functions for deep neural networks},
  author={Szanda{\l}a, Tomasz},
  booktitle={Bio-inspired neurocomputing},
  pages={203--224},
  year={2020},
  publisher={Springer}
}

@article{beg2021ubermag,
  title={Ubermag: Toward more effective micromagnetic workflows},
  author={Beg, Marijan and Lang, Martin and Fangohr, Hans},
  journal={IEEE Transactions on Magnetics},
  volume={58},
  number={2},
  pages={1--5},
  year={2021},
  publisher={IEEE}
}

@article{polley2022progress,
  title={Progress toward picosecond on-chip magnetic memory},
  author={Polley, Debanjan and Pattabi, Akshay and Chatterjee, Jyotirmoy and Mondal, Sucheta and Jhuria, Kaushalya and Singh, Hanuman and Gorchon, Jon and Bokor, Jeffrey},
  journal={Applied Physics Letters},
  volume={120},
  number={14},
  year={2022},
  publisher={AIP Publishing}
}

@article{polley2023picosecond,
  title={Picosecond spin-orbit torque--induced coherent magnetization switching in a ferromagnet},
  author={Polley, Debanjan and Pattabi, Akshay and Rastogi, Ashwin and Jhuria, Kaushalya and Diaz, Eva and Singh, Hanuman and Lemaitre, Aristide and Hehn, Michel and Gorchon, Jon and Bokor, Jeffrey},
  journal={Science Advances},
  volume={9},
  number={36},
  pages={eadh5562},
  year={2023},
  publisher={American Association for the Advancement of Science}
}

@article{mondal2023single,
title = {Single-shot switching in Tb/Co-multilayer based nanoscale magnetic tunnel junctions},
journal = {Journal of Magnetism and Magnetic Materials},
volume = {581},
pages = {170960},
year = {2023},
issn = {0304-8853},
doi = {https://doi.org/10.1016/j.jmmm.2023.170960},
url = {https://www.sciencedirect.com/science/article/pii/S0304885323006108},
author = {Sucheta Mondal and Debanjan Polley and Akshay Pattabi and Jyotirmoy Chatterjee and David Salomoni and Luis Aviles-Felix and Aurélien Olivier and Miguel Rubio-Roy and Bernard Diény and Liliana Daniela Buda Prejbeanu and Ricardo Sousa and Ioan Lucian Prejbeanu and Jeffrey Bokor}
}

@book{hebb-organization-of-behavior-1949,
	title = {The organization of behavior: {A} neuropsychological
		theory},
	author = {Hebb, Donald O.},
	publisher = Wiley,
	year = 1949
}

@article{hong2020machine,
	title={Machine learning and artificial neural network accelerated computational discoveries in materials science},
	author={Hong, Yang and Hou, Bo and Jiang, Hengle and Zhang, Jingchao},
	journal={Wiley Interdisciplinary Reviews: Computational Molecular Science},
	volume={10},
	number={3},
	pages={e1450},
	year={2020},
	publisher={Wiley Online Library}
}

@article{lee2016prediction,
	title={Prediction model of band gap for inorganic compounds by combination of density functional theory calculations and machine learning techniques},
	author={Lee, Joohwi and Seko, Atsuto and Shitara, Kazuki and Nakayama, Keita and Tanaka, Isao},
	journal={Physical Review B},
	volume={93},
	number={11},
	pages={115104},
	year={2016},
	publisher={APS}
}

@article{tarbi2022bandgap,
	title={Bandgap energy modeling of the deformed ternary GaAs1-uNu by artificial neural networks},
	author={Tarbi, A and Chtouki, T and Elkouari, Y and Erguig, H and Migalska-Zalas, A and Aissat, A},
	journal={Heliyon},
	volume={8},
	number={8},
	year={2022},
	publisher={Elsevier}
}

@article{agatonovic2000basic,
	title={Basic concepts of artificial neural network (ANN) modeling and its application in pharmaceutical research},
	author={Agatonovic-Kustrin, S and Beresford, Rosemary},
	journal={Journal of pharmaceutical and biomedical analysis},
	volume={22},
	number={5},
	pages={717--727},
	year={2000},
	publisher={Elsevier}
}

@article{baskin2016renaissance,
	title={A renaissance of neural networks in drug discovery},
	author={Baskin, Igor I and Winkler, David and Tetko, Igor V},
	journal={Expert opinion on drug discovery},
	volume={11},
	number={8},
	pages={785--795},
	year={2016},
	publisher={Taylor \& Francis}
}

@article{krizhevsky2012imagenet,
	title={Imagenet classification with deep convolutional neural networks},
	author={Krizhevsky, Alex and Sutskever, Ilya and Hinton, Geoffrey E},
	journal={Advances in neural information processing systems},
	volume={25},
	year={2012}
}

@article{gopalakrishnan2017deep,
	title={Deep convolutional neural networks with transfer learning for computer vision-based data-driven pavement distress detection},
	author={Gopalakrishnan, Kasthurirangan and Khaitan, Siddhartha K and Choudhary, Alok and Agrawal, Ankit},
	journal={Construction and building materials},
	volume={157},
	pages={322--330},
	year={2017},
	publisher={Elsevier}
}

@inproceedings{li2016evaluating,
	title={Evaluating the energy efficiency of deep convolutional neural networks on CPUs and GPUs},
	author={Li, Da and Chen, Xinbo and Becchi, Michela and Zong, Ziliang},
	booktitle={2016 IEEE international conferences on big data and cloud computing (BDCloud), social computing and networking (SocialCom), sustainable computing and communications (SustainCom)(BDCloud-SocialCom-SustainCom)},
	pages={477--484},
	year={2016},
	organization={IEEE}
}

@inproceedings{xiao2024convolutional,
	title={Convolutional neural network classification of cancer cytopathology images: taking breast cancer as an example},
	author={Xiao, MingXuan and Li, Yufeng and Yan, Xu and Gao, Min and Wang, Weimin},
	booktitle={Proceedings of the 2024 7th International Conference on Machine Vision and Applications},
	pages={145--149},
	year={2024}
}

@inproceedings{pan2015brain,
	title={Brain tumor grading based on neural networks and convolutional neural networks},
	author={Pan, Yuehao and Huang, Weimin and Lin, Zhiping and Zhu, Wanzheng and Zhou, Jiayin and Wong, Jocelyn and Ding, Zhongxiang},
	booktitle={2015 37th annual international conference of the IEEE engineering in medicine and biology society (EMBC)},
	pages={699--702},
	year={2015},
	organization={IEEE}
}

@article{yu2018neuro,
	title={Neuro-inspired computing with emerging nonvolatile memorys},
	author={Yu, Shimeng},
	journal={Proceedings of the IEEE},
	volume={106},
	number={2},
	pages={260--285},
	year={2018},
	publisher={IEEE}
}

@article{yu2021rram,
	title={RRAM for compute-in-memory: From inference to training},
	author={Yu, Shimeng and Shim, Wonbo and Peng, Xiaochen and Luo, Yandong},
	journal={IEEE Transactions on Circuits and Systems I: Regular Papers},
	volume={68},
	number={7},
	pages={2753--2765},
	year={2021},
	publisher={IEEE}
}

@article{baroni2022low,
	title={Low conductance state drift characterization and mitigation in resistive switching memories (RRAM) for artificial neural networks},
	author={Baroni, Andrea and Glukhov, Artem and Perez, Eduardo and Wenger, Christian and Ielmini, Daniele and Olivo, Piero and Zambelli, Cristian},
	journal={IEEE Transactions on Device and Materials Reliability},
	volume={22},
	number={3},
	pages={340--347},
	year={2022},
	publisher={IEEE}
}

@article{burr2015experimental,
	title={Experimental demonstration and tolerancing of a large-scale neural network (165 000 synapses) using phase-change memory as the synaptic weight element},
	author={Burr, Geoffrey W and Shelby, Robert M and Sidler, Severin and Di Nolfo, Carmelo and Jang, Junwoo and Boybat, Irem and Shenoy, Rohit S and Narayanan, Pritish and Virwani, Kumar and Giacometti, Emanuele U and others},
	journal={IEEE Transactions on Electron Devices},
	volume={62},
	number={11},
	pages={3498--3507},
	year={2015},
	publisher={IEEE}
}

@article{burr2016recent,
	title={Recent progress in phase-change memory technology},
	author={Burr, Geoffrey W and Brightsky, Matthew J and Sebastian, Abu and Cheng, Huai-Yu and Wu, Jau-Yi and Kim, Sangbum and Sosa, Norma E and Papandreou, Nikolaos and Lung, Hsiang-Lan and Pozidis, Haralampos and others},
	journal={IEEE Journal on Emerging and Selected Topics in Circuits and Systems},
	volume={6},
	number={2},
	pages={146--162},
	year={2016},
	publisher={IEEE}
}

@article{sebastian2019computational,
	title={Computational phase-change memory: Beyond von Neumann computing},
	author={Sebastian, Abu and Le Gallo, Manuel and Eleftheriou, Evangelos},
	journal={Journal of Physics D: Applied Physics},
	volume={52},
	number={44},
	pages={443002},
	year={2019},
	publisher={IOP Publishing}
}

@article{wang2024reconfigurable,
	title={Reconfigurable Multilevel Storage and Neuromorphic Computing Based on Multilayer Phase-Change Memory},
	author={Wang, Lu and Ma, Ge and Yan, Senhao and Cheng, Xiaomin and Miao, Xiangshui},
	journal={ACS Applied Materials \& Interfaces},
	volume={16},
	number={40},
	pages={54829--54836},
	year={2024},
	publisher={ACS Publications}
}

@article{sengupta2016proposal,
	title={Proposal for an all-spin artificial neural network: Emulating neural and synaptic functionalities through domain wall motion in ferromagnets},
	author={Sengupta, Abhronil and Shim, Yong and Roy, Kaushik},
	journal={IEEE transactions on biomedical circuits and systems},
	volume={10},
	number={6},
	pages={1152--1160},
	year={2016},
	publisher={IEEE}
}

@article{ross2023multilayer,
	title={Multilayer spintronic neural networks with radiofrequency connections},
	author={Ross, Andrew and Leroux, Nathan and De Riz, Arnaud and Markovi{\'c}, Danijela and Sanz-Hern{\'a}ndez, D{\'e}dalo and Trastoy, Juan and Bortolotti, Paolo and Querlioz, Damien and Martins, Leandro and Benetti, Luana and others},
	journal={Nature Nanotechnology},
	volume={18},
	number={11},
	pages={1273--1280},
	year={2023},
	publisher={Nature Publishing Group UK London}
}

@article{chen2023spintronic,
	title={Spintronic devices for high-density memory and neuromorphic computing--A review},
	author={Chen, BingJin and Zeng, Minggang and Khoo, Khoong Hong and Das, Debasis and Fong, Xuanyao and Fukami, Shunsuke and Li, Sai and Zhao, Weisheng and Parkin, Stuart SP and Piramanayagam, SN and others},
	journal={Materials Today},
	volume={70},
	pages={193--217},
	year={2023},
	publisher={Elsevier}
}

@article{kumar2024bimodal,
	title={Bimodal alteration of cognitive accuracy for spintronic artificial neural networks},
	author={Kumar, Anuj and Das, Debasis and Lin, Dennis JX and Huang, Lisen and Yap, Sherry LK and Tan, Hang Khume and Lim, Royston JJ and Tan, Hui Ru and Toh, Yeow Teck and Ter Lim, Sze and others},
	journal={Nanoscale Horizons},
	volume={9},
	number={9},
	pages={1522--1531},
	year={2024},
	publisher={Royal Society of Chemistry}
}

@article{seifert2016efficient,
	title={Efficient metallic spintronic emitters of ultrabroadband terahertz radiation},
	author={Seifert, Tom and Jaiswal, S and Martens, U and Hannegan, J and Braun, Lukas and Maldonado, Pablo and Freimuth, F and Kronenberg, A and Henrizi, J and Radu, I and others},
	journal={Nature photonics},
	volume={10},
	number={7},
	pages={483--488},
	year={2016},
	publisher={Nature Publishing Group UK London}
}

@article{fan2015stt,
	title={STT-SNN: A spin-transfer-torque based soft-limiting non-linear neuron for low-power artificial neural networks},
	author={Fan, Deliang and Shim, Yong and Raghunathan, Anand and Roy, Kaushik},
	journal={IEEE Transactions on Nanotechnology},
	volume={14},
	number={6},
	pages={1013--1023},
	year={2015},
	publisher={IEEE}
}

@article{finocchio2021promise,
	title={The promise of spintronics for unconventional computing},
	author={Finocchio, Giovanni and Di Ventra, Massimiliano and Camsari, Kerem Y and Everschor-Sitte, Karin and Amiri, Pedram Khalili and Zeng, Zhongming},
	journal={Journal of Magnetism and Magnetic Materials},
	volume={521},
	pages={167506},
	year={2021},
	publisher={Elsevier}
}

@inproceedings{zhang2014spintronics,
	title={Spintronics for low-power computing},
	author={Zhang, Yue and Zhao, Weisheng and Klein, Jacques-Olivier and Kang, Wang and Querlioz, Damien and Zhang, Youguang and Ravelosona, Dafin{\'e} and Chappert, Claude},
	booktitle={2014 Design, Automation \& Test in Europe Conference \& Exhibition (DATE)},
	pages={1--6},
	year={2014},
	organization={IEEE}
}

@article{fong2015spin,
	title={Spin-transfer torque devices for logic and memory: Prospects and perspectives},
	author={Fong, Xuanyao and Kim, Yusung and Yogendra, Karthik and Fan, Deliang and Sengupta, Abhronil and Raghunathan, Anand and Roy, Kaushik},
	journal={IEEE Transactions on Computer-Aided Design of Integrated Circuits and Systems},
	volume={35},
	number={1},
	pages={1--22},
	year={2015},
	publisher={IEEE}
}

@article{parkin2004giant,
	title={Giant tunnelling magnetoresistance at room temperature with MgO (100) tunnel barriers},
	author={Parkin, Stuart SP and Kaiser, Christian and Panchula, Alex and Rice, Philip M and Hughes, Brian and Samant, Mahesh and Yang, See-Hun},
	journal={Nature materials},
	volume={3},
	number={12},
	pages={862--867},
	year={2004},
	publisher={Nature Publishing Group UK London}
}

@article{fong2011bit,
	title={Bit-cell level optimization for non-volatile memories using magnetic tunnel junctions and spin-transfer torque switching},
	author={Fong, Xuanyao and Choday, Sri Harsha and Roy, Kaushik},
	journal={IEEE Transactions on Nanotechnology},
	volume={11},
	number={1},
	pages={172--181},
	year={2011},
	publisher={IEEE}
}

@article{wang2018novel,
	title={A novel MTJ-based non-volatile ternary content-addressable memory for high-speed, low-power, and high-reliable search operation},
	author={Wang, Chengzhi and Zhang, Deming and Zeng, Lang and Deng, Erya and Chen, Jie and Zhao, Weisheng},
	journal={IEEE Transactions on Circuits and Systems I: Regular Papers},
	volume={66},
	number={4},
	pages={1454--1464},
	year={2018},
	publisher={IEEE}
}

@inproceedings{he2017exploring,
	title={Exploring STT-MRAM based in-memory computing paradigm with application of image edge extraction},
	author={He, Zhezhi and Angizi, Shaahin and Fan, Deliang},
	booktitle={2017 IEEE International Conference on Computer Design (ICCD)},
	pages={439--446},
	year={2017},
	organization={IEEE}
}

@article{cui2020maximized,
	title={Maximized lateral inhibition in paired magnetic domain wall racetracks for neuromorphic computing},
	author={Cui, Can and Akinola, Otitoaleke G and Hassan, Naimul and Bennett, Christopher H and Marinella, Matthew J and Friedman, Joseph S and Incorvia, Jean Anne C},
	journal={Nanotechnology},
	volume={31},
	number={29},
	pages={294001},
	year={2020},
	publisher={IOP Publishing}
}

@article{das2023bilayer,
	title={Bilayer-skyrmion-based design of neuron and synapse for spiking neural network},
	author={Das, Debasis and Cen, Yunuo and Wang, Jianze and Fong, Xuanyao},
	journal={Physical Review Applied},
	volume={19},
	number={2},
	pages={024063},
	year={2023},
	publisher={APS}
}

@misc {OOMMF,
	title = {{OOMMF} User's Guide, Version 2.0a0},
	note = {http://math.nist.gov/oommf/},
	year = {2019},
	author = {Donahue , Michael Joseph and Porter , Donald Gene},
	url = {https://math.nist.gov/oommf/}
}

@MISC {DMI_extension,
	author = "Rohart, Stanislas and Thiaville, Andre",
	title = "{DMExchange6Ngbr: OOMMF} Oxs Extension Modules",
	note = "https://math.nist.gov/oommf/contrib/oxsext",
	year = "2012",
	url = {https://math.nist.gov/oommf/contrib/oxsext/},
	
}

@article{huang2017magnetic,
	title={Magnetic skyrmion-based synaptic devices},
	author={Huang, Yangqi and Kang, Wang and Zhang, Xichao and Zhou, Yan and Zhao, Weisheng},
	journal={Nanotechnology},
	volume={28},
	number={8},
	pages={08LT02},
	year={2017},
	publisher={IOP Publishing}
}

@article{ababei2021neuromorphic,
	title={Neuromorphic computation with a single magnetic domain wall},
	author={Ababei, Razvan V and Ellis, Matthew OA and Vidamour, Ian T and Devadasan, Dhilan S and Allwood, Dan A and Vasilaki, Eleni and Hayward, Thomas J},
	journal={Scientific Reports},
	volume={11},
	number={1},
	pages={15587},
	year={2021},
	publisher={Nature Publishing Group UK London}
}

@article{noh2012effects,
	title={Effects of notch shape on the magnetic domain wall motion in nanowires with in-plane or perpendicular magnetic anisotropy},
	author={Noh, Su Jung and Miyamoto, Yasuyoshi and Okuda, Mitsunobu and Hayashi, Naoto and Keun Kim, Young},
	journal={Journal of Applied Physics},
	volume={111},
	number={7},
	year={2012},
	publisher={AIP Publishing}
}

@article{wang2023spintronic,
	title={Spintronic leaky-integrate-fire spiking neurons with self-reset and winner-takes-all for neuromorphic computing},
	author={Wang, Di and Tang, Ruifeng and Lin, Huai and Liu, Long and Xu, Nuo and Sun, Yan and Zhao, Xuefeng and Wang, Ziwei and Wang, Dandan and Mai, Zhihong and others},
	journal={Nature Communications},
	volume={14},
	number={1},
	pages={1068},
	year={2023},
	publisher={Nature Publishing Group UK London}
}

@article{liu2024domain,
	title={Domain wall magnetic tunnel junction-based artificial synapses and neurons for all-spin neuromorphic hardware},
	author={Liu, Long and Wang, Di and Wang, Dandan and Sun, Yan and Lin, Huai and Gong, Xiliang and Zhang, Yifan and Tang, Ruifeng and Mai, Zhihong and Hou, Zhipeng and others},
	journal={Nature Communications},
	volume={15},
	number={1},
	pages={4534},
	year={2024},
	publisher={Nature Publishing Group UK London}
}

@article{yang2021integrated,
	title={Integrated neuromorphic computing networks by artificial spin synapses and spin neurons},
	author={Yang, Seungmo and Shin, Jeonghun and Kim, Taeyoon and Moon, Kyoung-Woong and Kim, Jaewook and Jang, Gabriel and Hyeon, Da Seul and Yang, Jungyup and Hwang, Chanyong and Jeong, YeonJoo and others},
	journal={NPG Asia Materials},
	volume={13},
	number={1},
	pages={11},
	year={2021},
	publisher={Nature Publishing Group UK London}
}

@article{lone2025spintronic,
	title={Spintronic memtransistor leaky integrate and fire neuron for spiking neural networks},
	author={Lone, Aijaz H and Tang, Meng and Rahimi, Daniel N and Zou, Xuecui and Zheng, Dongxing and Fariborzi, Hossein and Zhang, Xixiang and Setti, Gianluca},
	journal={Advanced Electronic Materials},
	volume={11},
	number={13},
	pages={2500091},
	year={2025},
	publisher={Wiley Online Library}
}

@article{jung2022crossbar,
	title={A crossbar array of magnetoresistive memory devices for in-memory computing},
	author={Jung, Seungchul and Lee, Hyungwoo and Myung, Sungmeen and Kim, Hyunsoo and Yoon, Seung Keun and Kwon, Soon-Wan and Ju, Yongmin and Kim, Minje and Yi, Wooseok and Han, Shinhee and others},
	journal={Nature},
	volume={601},
	number={7892},
	pages={211--216},
	year={2022},
	publisher={Nature Publishing Group UK London}
}

@article{raj2026reconfigurable,
	title={Reconfigurable Bio-Plausible Spiking Neurons Based on Antiferromagnetic Skyrmions Utilizing Anisotropy Gradient},
	author={Raj, Ravish Kumar and Jony, Sharuar Hossain and Sengupta, Purbak and Shreya, Sonal},
	journal={IEEE Magnetics Letters},
	year={2026},
	publisher={IEEE}
}

@article{thiaville2012dynamics,
	title={Dynamics of Dzyaloshinskii domain walls in ultrathin magnetic films},
	author={Thiaville, Andr{\'e} and Rohart, Stanislas and Ju{\'e}, {\'E}milie and Cros, Vincent and Fert, Albert},
	journal={Europhysics Letters},
	volume={100},
	number={5},
	pages={57002},
	year={2012},
	publisher={EDP Sciences, IOP Publishing and Societ{\`a} Italiana di Fisica}
}

@article{emori2013current,
	title={Current-driven dynamics of chiral ferromagnetic domain walls},
	author={Emori, Satoru and Bauer, Uwe and Ahn, Sung-Min and Martinez, Eduardo and Beach, Geoffrey SD},
	journal={Nature materials},
	volume={12},
	number={7},
	pages={611--616},
	year={2013},
	publisher={Nature Publishing Group UK London}
}

@article{shreya2023granular,
	title={Granular vortex spin-torque nano oscillator for reservoir computing},
	author={Shreya, Sonal and Jenkins, AS and Rezaeiyan, Yasser and Li, Ren and B{\"o}hnert, T and Benetti, Luana and Ferreira, Ricardo and Moradi, Farshad and Farkhani, Hooman},
	journal={Scientific Reports},
	volume={13},
	number={1},
	pages={16722},
	year={2023},
	publisher={Nature Publishing Group UK London}
}
	
\end{document}